%% file: main.tex
\documentclass[conference,compsoc]{IEEEtran}
\usepackage[nocompress]{cite}
\usepackage{graphicx}
\usepackage{amsmath,amssymb}
\usepackage{url}
\input{packages.tex}
\input{abbrv.tex}

\begin{document}

\title{Consumer Beware!\\
Exploring Data Brokers' CCPA Compliance}


\author{\IEEEauthorblockN{Elina van Kempen, Isita Bagayatkar, Pavel Frolikov, Chloe Georgiou, and Gene Tsudik}
\IEEEauthorblockA{University of California, Irvine\\
\{evankemp, ibagayat, pavel, cgeorgio, gtsudik\}@uci.edu}}

\maketitle
\begin{abstract}
Data brokers collect and sell the personal information of  millions of individuals, 
often without their knowledge or consent. The California Consumer Privacy Act (CCPA) 
grants consumers the legal right to request access to, or deletion of, their data. 
To facilitate these requests, California maintains an official registry of 
data brokers. However, the extent to which these entities comply with the law is unclear. 

This paper presents the first large-scale, systematic study of CCPA compliance of
all $\bm{543}$ officially registered data brokers. Data access requests were manually submitted 
to each broker, followed by in-depth analyses of their responses (or lack thereof). 
Above 40\% failed to respond at all, in an apparent violation of the CCPA. 
Data brokers that responded requested personal information as part of their identity verification 
process, including details they had not previously collected. Paradoxically, this means that 
exercising one's privacy 
rights under CCPA introduces new privacy risks. 

Our findings reveal rampant non-compliance and lack of standardization of the
data access request process. These issues highlight an urgent need for stronger enforcement,  
clearer guidelines, and standardized, periodic compliance checks to enhance consumers'
privacy protections and improve data broker accountability.

\end{abstract}
\IEEEpeerreviewmaketitle


\input{00content}

\bibliographystyle{IEEEtran}
\bibliography{references}

\input{appendix}

\end{document}

%% file: packages.tex
\usepackage{multirow}
\usepackage{caption}
\usepackage{subcaption}
\usepackage{arydshln}
\usepackage{paralist}
\usepackage{enumitem}
\usepackage{xspace}
\usepackage{adjustbox}
\usepackage{color} 
\usepackage[T1]{fontenc}
\definecolor{commentorange}{RGB}{255,127,0}
\usepackage{bm}

%% file: abbrv.tex
\newcommand{\dbr}{{\ensuremath{\sf{\mathcal{DBR}}}}\xspace}
\newcommand{\DBR}{\dbr}
\newcommand{\dbrs}{{\ensuremath{\sf{\mathcal{DBR}s}}}\xspace}
\newcommand{\dbrspossessive}{{\ensuremath{\sf{\mathcal{DBR}s}}'}\xspace}

\newcommand{\DBRsBOLD}{{$\bm{\mathcal{DBR}\mathsf{s}}$}\xspace}

\newcommand{\new}[1]{\textcolor{black}{#1}}

%% file: 00content.tex
\section{Introduction}\label{sec:intro}
Data brokers (\dbrs) operate largely hidden from public view: collecting, aggregating, and 
selling personal information (PI) of consumers without their knowledge or consent. 
These entities systematically harvest data from various sources: public records, online activities, 
social media profiles, and even other \dbrs. They routinely analyze this collected data to 
determine sensitive information such as purchasing behavior, financial status, and  
health conditions. Then, \dbrs monetize this information by selling it to various third parties, 
including companies, government agencies, and individuals. These sales are often completed 
without informing the consumer. One notorious and all-too-familiar class of \dbrs 
corresponds to so-called ``people search'' websites: online platforms that aggregate and 
market individual consumers' PI. Often, some PI is available for free, and additional information 
is available for gradually increasing prices. Usually, there is a prominent
pitch to pay a fee in order to see detailed PI, with variable pricing schemes.
Thus, any malicious actor can gain access to, and misuse, consumer PI to mount
identity theft, fraud, or phishing attacks. 

\dbrs operate without current relationships with consumers whose data they process, 
unlike traditional data aggregators (such as credit reporting agencies) with which consumers 
knowingly share their data to obtain presumably useful services. 
This fundamental difference results in consumers being unaware that their PI
is being collected, analyzed, and made available for sale. Actual implications of such 
unmitigated access to consumers' PI remain poorly understood by the general public. 

To remedy the situation, several jurisdictions enacted data protection laws, e.g., General 
Data Protection Regulation (GDPR)\cite{GDPR} in the European Union, California Consumer 
Privacy Act (CCPA)\cite{CCPA} in California, and Lei Geral de Proteção de Dados (LGPD)\cite{LGPD} 
in Brazil. Though specific provisions differ among them, these laws aim to give individuals 
greater control over their PI. Notably, they grant consumers the rights to access, correct, 
and delete PI held by either businesses or other organizations. 
Under CCPA, consumers exercise these rights by making a \emph{Verifiable Consumer Request (VCR)}:
\begin{quote}
"a request made by a consumer ..., and that the business can verify, using commercially reasonable 
methods, ... to be the consumer about whom the business has collected personal information"
(1798.140(ak)) \cite{CCPA}.
\end{quote}
The California Privacy Protection Agency (CPPA) specifically targets \dbrs. Besides CCPA 
compliance, \dbrs must abide by the Data Broker Registration law, which requires each \dbr 
to register annually with the CPPA. The resulting registry is made publicly available 
on the CPPA website\cite{california_data_broker_registry}. 

Composing and submitting a VCR is burdensome for consumers because there is no 
standard process for doing so. Each \dbr has its own method for consumers to submit a 
VCR, such as by filling out a form on the website, sending an email, making a phone-call,
or even having to go through a multi-step process. The necessary steps are typically 
(supposed to be) contained within a \dbrspossessive privacy policy. 

Identity verification adds another layer of complexity. A \dbr must verify the requesting consumer's 
identity before releasing the data in order to prevent data breaches. However, this verification process 
is not standardized and is taxing for the average consumer.

Accessing consumer data is especially challenging when interacting with \dbrs. Unlike 
other business entities, which users expect to collect PI, there is no way of knowing 
beforehand whether a \dbr collected PI of a particular consumer. Thus, the consumer is 
forced to \emph{blindly} make VCRs to a long list of \dbrs. 
The identity verification process is also somewhat Kafkaesque: 
\begin{quote}
How can a consumer prove their identity to a \dbr that may, or may \emph{not}, have their PI?
\end{quote}
Throughout this paper, we distinguish between: (1) Personal Information (\textbf{PI}), i.e., any 
information a \dbr may have about a consumer, and (2) Personally Identifiable Information (\textbf{PII}), 
which is a subset of PI that specifically identifies a consumer, e.g., name, address, or Social Security Number (SSN). 
This distinction is important for understanding privacy implications of CCPA enforcement. 
For example, the consumer's device model is PI, and not PII, because many consumers share the same
device type; however, a driver’s license number is PII since it uniquely identifies an individual. 
As shown later, this requirement to provide PII for identity verification creates an unintended privacy 
paradox when exercising one's CCPA rights. 

\subsection{Research Questions}
Motivated by curiosity about the current state of CCPA compliance and the practical challenges consumers 
face when exercising their privacy rights, this work seeks to answer the 
following three research questions:
\begin{compactitem}
    \item [\textbf{RQ1.}] \textbf{
    \new{What challenges are encountered in the VCR submission process, especially in the identity verification step?}}
\end{compactitem}
\noindent VCR submission is not a streamlined process. \dbrs can 
simplify it by adopting user-friendly privacy policies and employing reasonable VCR submission 
methods. We measured the time needed for the researcher to find each \dbr's contact information 
on their privacy policy, as well as the time needed to submit a VCR, along with the amount 
of PII needed for the \dbr to verify the researcher's identity. 
\begin{compactitem}
    \item[\textbf{RQ2.}] \textbf{To what extent do \DBRsBOLD comply with CCPA after VCR submission? }
\end{compactitem}
\noindent The goal is to discover what fraction of registered \dbrs comply,
fully or in part, with CCPA requirements described in Section \ref{sec:background}. 
After measuring VCR response rates and response timelines, we consider the factors that 
correlate with CCPA compliance. 

\begin{compactitem}
    \item[\textbf{RQ3.}] \textbf{What kind of data do \DBRsBOLD collect and how is it shared with the 
    consumer following a VCR?}
\end{compactitem}
\noindent After receiving the PI from \dbrs, 
we classify both the type of PI collected and its accuracy. We consider whether this 
data is sent securely to consumers, and whether it is provided in a standard and usable format 
(e.g., TXT, CSV, PDF, or JSON) as required by CCPA, rather than in some proprietary, 
obscure, or otherwise unparsable format. That PI must also be provided "...in a format that is 
easily understandable for the average consumer" (1798.130 (a)(3)(B)(iii)) \cite{CCPA}.

\subsection{Study Overview}
This paper reports on a comprehensive and systematic study of the \dbr ecosystem.
The study involved all $543$ \dbrs duly registered in California, listed under 
the California Privacy Protection Agency.
We examined six key aspects of the VCR submission process: (1) consumer burden of composing VCRs, 
(2) variability in identity verification, (3) response time fluctuations, 
(4) quality of responses (i.e., content), (5) what actual PI is being collected, and 
(6) privacy issues, if any, that arise in the process of requesting PI.

To begin the study, we submitted VCRs to all $543$ \dbrs according to their individual
guidelines, if those were available. We then measured their
response timeline and analyzed common practices. We found that only 57\% responded 
to VCRs, meaning that 43\% are blatantly non-compliant.

We found that the majority of \dbrs did not have any PI about the researcher who submitted VCRs.
However, we still had to send PII to every \dbr as 
part of their identity verification process. 

Although some prior work studied CCPA and/or GDPR compliance, its focus was on entities 
with which the consumer has a direct relationship 
\cite{DBLP:conf/soups/MartinoRWQLA19,DBLP:journals/popets/MartinoMQAL22,adhatarao2021ip,
DBLP:conf/icws/BufalieriMMS20,DBLP:conf/apf/BonifaceFBLS19,wong2019right,DBLP:conf/sicherheit/HerrmannL16,
DBLP:conf/IEEEares/KrogerLH20,DBLP:journals/popets/SamarinKSBYWAFHE23}. 
Furthermore, previous research on \dbrs examined 
only a small subset of \dbrs: people search websites \cite{DBLP:journals/popets/TakeYBGFMG24}. 
Other types of \dbrs have not been investigated.
A previous study evaluated the usability of data access and removal of 20 such websites \cite{DBLP:journals/popets/TakeYBGFMG24}, 
offering a rather narrow view of the \dbr ecosystem. 

\subsection{Organization} 
Section \ref{sec:relwork} presents an overview of prior work, and Section 
\ref{sec:background} summarizes relevant 
CCPA sections. The methodology is described in Section \ref{sec:meth}, including 
ethical considerations in \ref{sec:meth:eth}. Section \ref{sec:results} follows with results. 
Finally, privacy issues and study outcomes are described in Section \ref{sec:disc}.


\section{Related Work}\label{sec:relwork}
Related work is summarized in Table \ref{tab:relwork}, which specifies the number of entities
to which VCRs could be submitted, followed by the total number of entities
involved in each study, in parentheses. Note that, for the sake of simplicity and uniformity,
we use the term {\bf VCR} to refer to privacy rights requests, and the term \textbf{consumer} 
to refer to an individual whose PI is collected, regardless of the specific 
legal framework in the context of which they occur.

Most prior work on compliance with privacy laws (in terms of data access VCRs)
focused on studying popular websites or apps 
\cite{DBLP:conf/soups/MartinoRWQLA19,DBLP:journals/popets/MartinoMQAL22,
adhatarao2021ip,DBLP:conf/icws/BufalieriMMS20,DBLP:conf/apf/BonifaceFBLS19,
wong2019right,DBLP:conf/sicherheit/HerrmannL16,DBLP:conf/IEEEares/KrogerLH20,
DBLP:journals/popets/SamarinKSBYWAFHE23}. 
Some prior results \cite{DBLP:conf/icws/BufalieriMMS20,DBLP:conf/sicherheit/HerrmannL16,DBLP:conf/IEEEares/KrogerLH20,DBLP:journals/popets/SamarinKSBYWAFHE23} 
ensured that targeted entities already had data about the specific consumer. To do so, 
they picked popular websites or applications, created accounts 
whenever possible, and used the services for a certain time. 
For example, \cite{DBLP:conf/soups/MartinoRWQLA19,DBLP:journals/popets/MartinoMQAL22,wong2019right}, 
focused on selected entities that had already collected data about the researcher(s). 
Also, some prior work  \cite{DBLP:conf/soups/MartinoRWQLA19,DBLP:journals/popets/MartinoMQAL22} also
examined vulnerabilities in the data request process and tried to access consumer 
PI through impersonation.  

\new{
Borem et al. \cite{borem2024data} investigated data export procedures, in response to 
data access requests, for 5 major service providers: Amazon, Facebook, Google, Spotify, 
and Uber. A total of 33 participants requested data; each participant had $\geq~2$ years of use history with the 
relevant entity. Participants found that data exports were 
confusing and overwhelming, due to large amounts of information
and a lack of explanation for the types of data collected and/or the reasons for 
collection.
}

Adhatarao et al. \cite{adhatarao2021ip} reported on attempts to submit VCRs 
to websites they visited using only an IP address for identification, which was 
unsuccessful for all attempts.

Boniface et al. \cite{DBLP:conf/apf/BonifaceFBLS19} and Urban et al. 
\cite{DBLP:conf/esorics/UrbanTDHP19} evaluated submission of 
VCRs to well-known third-party tracking services. The former
did not report on the results of VCRs, instead focusing on security aspects 
of the requester's identity verification and the ease of making a request. The latter
used stored cookies created by tracking services for identification: only 58\% of
tracking services replied, whether positively or negatively. 

\new{Very recently, He et al. \cite{HePets25} analyzed pricing, coverage, and 
efficacy of 10 PII removal services, all of which claim to handle the VCR process 
with data brokers on behalf of consumers. Results showed that 
PII removal services differed greatly in terms of which data brokers they cover. 
They also generally failed to remove the majority of users' records from the 
data brokers.
}

Take et al. \cite{DBLP:journals/popets/TakeYBGFMG24} is the most similar to our work.
It reported on a study testing the compliance of 20 people-search websites, which is a 
type of \dbr, e.g., Intellus, Truthfinder, and Whitepages.
VCRs were made for each such website under both GDPR and CCPA, 
depending on the researchers' residence. Response data was compared to that received 
after paying for user reports. However, results did not specify whether
requests made under one law yielded better results than those made under another. 
Furthermore, the timeline of responses was not provided. Also, we believe that 
significant results or trends about the \dbr landscape cannot be derived from the 
qualitative analysis of only 20 people-search websites.

In contrast, the study described in this paper systematically and comprehensively
evaluates the behavior of \textbf{all} registered \dbrs after submission of CCPA data access VCRs.

\begin{table}[ht] 
    \centering
        \caption{Comparison of this work to relevant prior results.}
     \resizebox{0.95\linewidth}{!}{
    \begin{tabular}{lrccc}
    \hline
         & \textbf{\# \new{Entities}} & \textbf{Law} & \textbf{Resp. rate} & \textbf{Year} \\
         \hline
         \multicolumn{4}{l}{$\bm{\mathcal{DBR}\mathsf{s}}$}\\
             \textit{this} & 454 (543) & CCPA & 57\% & 2024-2025 \rule[-1.2ex]{0pt}{0pt}\\ \hdashline[2pt/2pt]   
        \multicolumn{4}{l}{{\rule{0pt}{3ex}People Search Websites}}\\
         Take et al. \cite{DBLP:journals/popets/TakeYBGFMG24} & 20 & CCPA, GDPR & 82-100\% & 2024  \\

        \hline
         \multicolumn{4}{l}{\textbf{Tracking services}}\\
         Boniface et al. \cite{DBLP:conf/apf/BonifaceFBLS19} & 25 (30) & GDPR & - & 2018-2019\\
         Urban et al. \cite{DBLP:conf/esorics/UrbanTDHP19} & 36 (39) & GDPR & 58\% & 2018 \\
         
         \hline
         \multicolumn{4}{l}{\textbf{Common websites}} \\
         Martino et al. \cite{DBLP:journals/popets/MartinoMQAL22} & 40 & GDPR & 93\% & 2021\\
         Adhatarao et al. \cite{adhatarao2021ip} & 109 (124)  & GDPR & 57\% & 2021\\
         Bufalier et al. \cite{DBLP:conf/icws/BufalieriMMS20} & 317 (341) & GDPR & 71\% & 2020 \\
         Martino et al. \cite{DBLP:conf/soups/MartinoRWQLA19} & 55 & GDPR & 93\% & 2019  \\
         Boniface et al. \cite{DBLP:conf/apf/BonifaceFBLS19} & 27 (50)  & {GDPR} & - & 2018-2019  \\
         Wong et al. \cite{wong2019right} & 229 (230) & GDPR & 75\% & 2018\\
         Herrmann et al. \cite{DBLP:conf/sicherheit/HerrmannL16} & 119 (120) & GDPR & 77\% & 2016\\
         \hline
         \multicolumn{4}{l}{\textbf{Smartphone apps}}\\
         Samarin et al. \cite{DBLP:journals/popets/SamarinKSBYWAFHE23} & 109 (160) & CCPA & 81\% & 2023 \\
         Kroger at al. \cite{DBLP:conf/IEEEares/KrogerLH20} & 216-224 (225) & GDPR & 77-83\% & 2015-2019 \\
         Herrmann et al. \cite{DBLP:conf/sicherheit/HerrmannL16} & 144 (150) & GDPR & 55\% & 2016\\

    \hline
    \end{tabular}}
    \label{tab:relwork}
\end{table}


\section{Background: CCPA and Data Broker Registration}\label{sec:background}
The General Data Protection Regulation (GDPR), enacted in 2016 by the European 
Union, established comprehensive privacy regulations and rights for European 
consumers \cite{GDPR}. It also inspired new privacy regulations internationally.
In particular, the California Consumer Privacy Act (CCPA), voted on in 2018 and 
amended in 2020 by the California Privacy Rights Act (CPRA), grants
California residents certain rights over the data collected about them by businesses 
\cite{CCPA}. These rights are commonly known as {\em the rights to know, correct, or 
delete PI collected by a business}. \\

\noindent \textbf{Personal information.} CCPA defines PI as 
information that:
\begin{quote}
    "identifies, relates to, describes, ..., or could reasonably be linked, 
    directly or indirectly, with a particular consumer or household." 
    (1798.140(v)) \cite{CCPA}
\end{quote}
This includes identifiers (names or usernames) and other consumer data, 
such as addresses, biometric information, geolocation information, and network activity.
It excludes publicly available information, i.e., public government records and 
consumer information made public by the consumer or another person, assuming no 
restrictions placed by the consumer.

In addition, CCPA distinguishes 
\emph{sensitive} PI, which includes an individual's SSN, driver's license or passport number, 
genetic information, racial or ethnic information, and account credentials. (1798.140(ae)) \cite{CCPA}. \\

\noindent $\bm{\mathcal{DBR}\mathsf{s}.}$ Businesses must comply with CCPA if they satisfy
one of the following (1798.140(d)) \cite{CCPA}:
\begin{compactenum}
    \item Have a gross annual revenue over \$25.625 million.
    \item Handle over 100,000 California residents' PI.
    \item Get more than 50\% of their revenue from the sale of California residents' 
    PI.
\end{compactenum}
The Data Broker Registration law establishes further 
registration requirements for \dbrs \cite{DataBrokerLaw}. A \dbr is defined as:
\begin{quote}
    "a business that knowingly collects and sells to third parties the personal 
    information of a consumer with whom the business does not have a direct relationship" 
    (1798.99.80(c)) \cite{DataBrokerLaw}. 
\end{quote}
As described in Section \ref{sec:intro}, unlike other businesses that collect 
data of their customers (i.e., consumers who previously used their services), \dbrs 
collect data from individuals who have never used their services. \dbrs specialize 
in collecting, analyzing, and monetizing consumer data by selling it to third parties. 

The Data Broker Registration law requires \dbrs to register annually\footnote{The California \dbr registry is 
available at \url{https://cppa.ca.gov/data_broker_registry/}} 
with the California Privacy Protection Agency (1798.99.82) \cite{DataBrokerLaw}. 
The registration fee is currently \$6,600, with a 2.99\% additional electronic 
payment fee (Tit. 11 Div. 6 \S 7600) \cite{CCR}. \dbrs that fail to register 
may face a fine of \$200 per day of non-compliance.
The law also requires \dbrs to undergo regular third-party audits starting in 
2028 (1798.99.86(c)) \cite{DataBrokerLaw}. \\

\noindent\textbf{VCR processing.}
CCPA encourages verification of a consumer's identity using an existing 
password-protected account, although it prohibits businesses from requiring a 
consumer to create an account in order to submit a VCR (1798.130(a)(2)(A)) \cite{CCPA}. 

CCPA provides guidelines for identity verification of 
non-account holders (Tit. 11 Div. 6 \S 7062) \cite{CCR}, and requires a 
business to verify the identity of the consumer to a:
\begin{compactitem}
    \item "reasonable degree of certainty", when a user requests to know categories
    of PI currently possessed by the business, by matching at least two "reliable" data points 
    provided by the consumer with data points possessed by the business.
    \item "reasonably {\underline{high}} degree of certainty", when a user requests to 
    know specific pieces of PI, by matching at least 
    three "reliable" data points provided by the consumer with data points 
    possessed by the business, along with a signed declaration under penalty of perjury.
\end{compactitem}
CCPA states that a business may not use the PII provided 
during the identity verification process for any other purposes 
(1798.130(a)(7)) \cite{CCPA}.

Once a business verifies the consumer's identity, it must: 
(1) confirm receipt of the request within 10 business days 
(Tit. 11 Div. 6 \S  7021(a)) \cite{CCR}, and (2) answer the request within 45
calendar days. If needed, a business may extend its response time by an 
additional 45 calendar days, as long as it notifies the consumer 
(1798.130(a)(2)(A)) \cite{CCPA}. CCPA requires businesses to deliver 
the information by mail or electronically, "in a readily usable format 
that allows the consumer to transmit this information from one entity to 
another entity without hindrance" (1798.130(a)(2)(A)) \cite{CCPA}.

\subsection{$\bm{\mathcal{DBR}\mathsf{s}}$ Location}
Almost all \dbrs are registered to an address within the United States, 
with only 
28 \dbrs registered abroad. Table \ref{tab:countries} specifies the 
geographical distribution of \dbrs to which VCRs could be submitted, by country. 

\begin{figure}[ht]
    \centering
    \includegraphics[width=\linewidth]{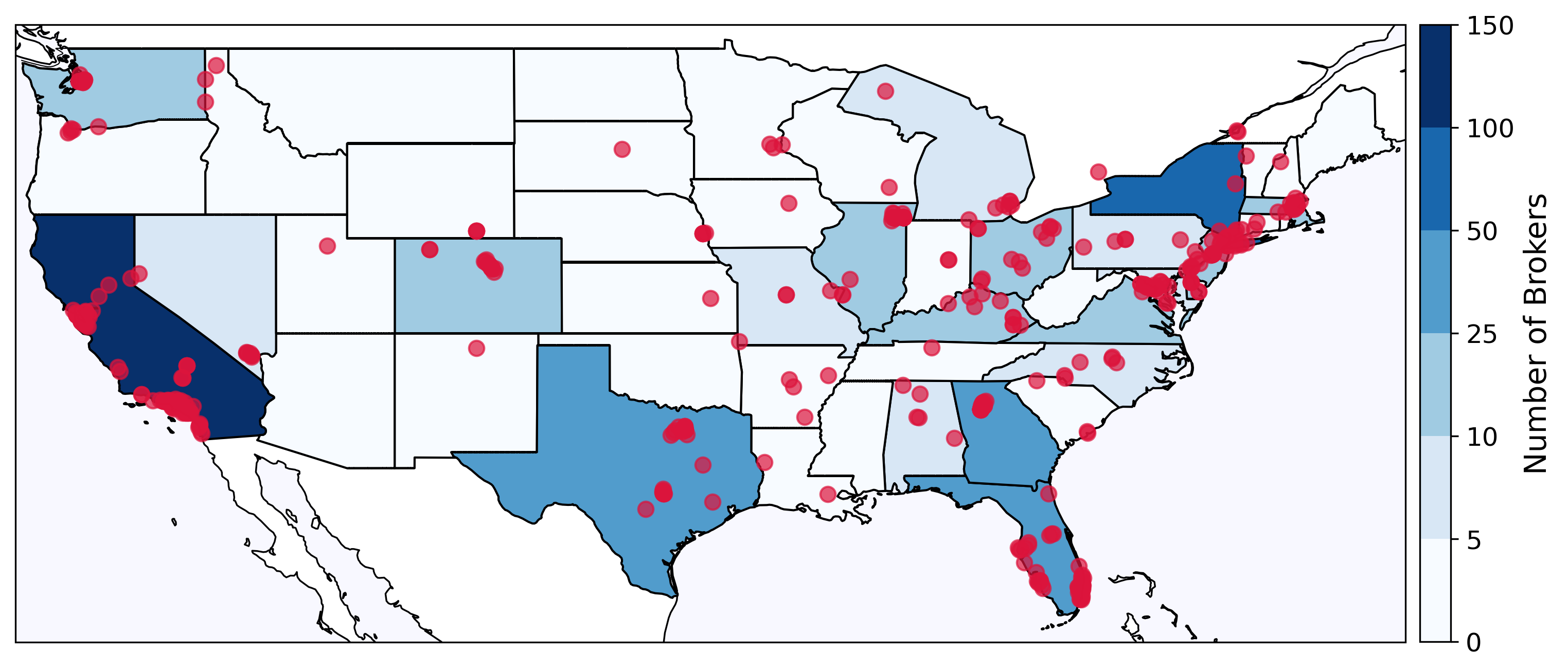}
    \caption{Map of US \dbrs.}
    \label{fig:map}
\end{figure}

\begin{table}[ht]
    \centering
        \caption{\dbrs by country of registration.
        }
    \label{tab:countries}
    \begin{tabular}{lr}
    \hline
      Country   &  \dbr Count\\
      \hline
      United States   &  515\\
      United Kingdom        &      10\\
Canada           &             5\\
Germany, Denmark           &            2 each\\
Malaysia, India, Poland, Sweden,         &             \\
Israel, New Zealand, Namibia, Italy, & \\
United Arab Emirates & 1 each\\
\hline
Total & 543 \\
\hline
    \end{tabular}
\end{table}
Within the United States, \dbrs are registered in 40 states.
A large number are registered in three states: California, New York, and Florida, 
with 
103, 
74, and 
46 registered \dbrs, respectively. 
Figure \ref{fig:map} shows a map of all \dbrs in the United States. Since no \dbrs are 
registered in Alaska or Hawaii, these states are not shown on the map.


\section{Methodology}\label{sec:meth}
%
Recall that our goal is to assess \dbrspossessive CCPA compliance. To this end, we submitted 
VCRs\footnote{From this point forward, VCR refers specifically to data access requests.} for PI 
collected by all registered \dbrs. We chose not to submit data 
deletion requests, since by doing so we would learn less information about \dbrspossessive 
data collection and response processes. In other words, a \dbr could simply reply
that relevant PI found in their database will be deleted, whether or not such data 
actually exists, with no proof of deletion.
Submitting a VCR guarantees a database lookup by the (compliant) \dbr,
which must then reply with collected PI or with a statement that it has none.

\subsection{Study Design}\label{sec:studes}
We investigated all $543$ registered \dbrs in the California Data Broker Registry during the 2024 year\cite{brokerlist} 
and submitted VCRs to exercise our "right to know/access". A single co-author (denoted "researcher" 
hereafter), California resident since 2021, submitted all VCRs to ensure consistency and privacy. \\

\noindent \textbf{Submitting VCRs.}
Each \dbr in the registry provided a URL to their website's homepage. After finding a privacy policy, 
the researcher located each \dbr's instructions for submitting VCRs. The researcher used each 
\dbr's preferred VCR submission method to provide \dbrs with the best circumstances and maximize 
the likelihood of getting a response. In situations where the \dbr-provided forms were 
non-functional, the researcher used an alternative email contact method listed in the 
privacy policy.  A standard (consistent) email template was used across all 
emailed VCRs; see in Appendix \ref{app:email}.

In a few cases, when neither a form-based nor an email-based contact method worked, 
the only option was making a phone-call.
Although CCPA requires \dbrs to provide at least a toll-free telephone number 
for submitting VCRs (1798.130(a)(1)(A))\cite{CCPA}, doing it by phone is not optimal 
since it leaves no proof or record of submission. Additionally, callbacks are easy to miss. 
Whenever a phone-call led to voicemail, the researcher left a message stating that 
they wanted to submit a VCR, along with their phone number for \dbrs to call back. 

To accurately track all VCRs and responses, a dedicated email address was created 
solely for the VCR submission purpose. If \dbrs requested alternative email addresses 
(e.g., a business email), the researcher provided them. 

The researcher duly provided all PII requested by \dbrs for identity verification purposes. 
If a copy of a government-issued ID was needed, the researcher sent a redacted version 
unless a full copy was specifically requested. \\

\noindent \textbf{Metrics.}
We measured VCR submission time for each \dbr: the timer started when the researcher landed 
on the \dbr's homepage and stopped when a form was submitted, an email was sent, or a phone-call ended. 
Thus, we measured the time needed to: (1) locate the privacy policy on the \dbr's website, (2) 
identify the VCR submission method within that policy, and (3) perform the actual VCR submission. 
The data collection period lasted roughly 7 months total, from late 2024 to early 2025. 

We recorded PII elements requested by each \dbr for VCR submission and identity verification.
To evaluate CCPA compliance, we tracked the response rate, timeline of responses, and the content of
responses. We recorded the dates when: (1) the original request was sent, (2) the request 
receipt acknowledgment was received, and (3) the response (if any) was received. If a \dbr 
requested additional PII, we recorded the time needed by the researcher to provide this 
information. In order to precisely and fairly assess response timeline, we exclude 
time taken by the researcher to submit the requested additional PII from our analysis. 
Thus, when a \dbr requests additional information, the \dbr is not penalized for the
time taken by the researcher to reply to the \dbr. \\

\noindent \textbf{Exclusions (Data Cleaning).}
$89$ out of $543$ registered \dbrs were excluded from the full study: \\

\noindent \textit{Duplicates.} $50$ corresponded to duplicate entries, i.e., different registry 
entries leading to the same website, email address, or form. We only submitted one request for each 
duplicate registration. \\

\noindent \textit{Noncompliant.} 
$8$ \dbrspossessive websites were completely inaccessible during the study period. 
$7$ \dbrs lacked any privacy policy or contact information to submit VCRs. 
$11$ \dbrs for which email-based VCRs could not be delivered. 
$2$ \dbrs only provided phone contacts, and there was no way to leave 
a voicemail; in each case, the researcher hung up after $5$ minutes of phone silence. 
Finally, 1 \dbr required notarization, although CCPA regulations (Tit. 11 Div. 6 \S  7060(e)) \cite{CCR} state that: 
"...a business may not require a consumer to provide a notarized affidavit to verify their 
identity unless the business compensates the consumer for the cost of notarization" \cite{CCR}. \\

\noindent \textit{Lack of Security.} 
2 \dbrs with non-HTTPS websites and forms were excluded for security and privacy reasons. \\

\noindent \textit{Unable to complete VCR.}
4 \dbrs required a website cookie ID, which was not present on the researcher's device(s).
3 \dbrs requested information that the researcher did not have, e.g.,
organizational position, website URL, or medical professional status. \\

\noindent \textit{"No longer a \dbr{}" (but registered anyway).} 1 \dbr announced 
on its main web page that it had exited consumer data sales. \\

\noindent After accounting for all exclusions, the analysis focused on the remaining 454 \dbrs, 
representing approximately 84$\%$ of the total.

\subsection{Ethical considerations}\label{sec:meth:eth}
The authors' Institutional Review Board officially declared that 
this study constitutes non-human subject research.
The lead researcher (using their own identity to compose VCRs) did so voluntarily, 
motivated primarily by the research goals. To preserve their privacy, 
only that researcher (who submitted all VCRs) had access to \dbr responses. The researcher 
could choose to withdraw from the study at any time and 
not submit VCRs to certain websites if they felt uncomfortable. In particular, as mentioned above, 
the researcher chose not to make VCRs to \dbrs that had non-HTTPS websites 
out of concern for their privacy. 

Also, recall that we observed that some \dbrs are registered more than once,
i.e., multiple \dbrspossessive registrations led to the same VCR submission form or 
email. We avoided contacting them multiple times to be respectful of their resources. If we 
accidentally contacted a \dbr twice, we only considered the first VCR for all further analysis.

Finally, besides the scientific goal of this study, the researcher had genuine and
legitimate interest in learning what PI about them was collected by \dbrs. 


\section{Results}\label{sec:results}
We now present quantitative results addressing \textbf{RQ1}, \textbf{RQ2}, and \textbf{RQ3}. 
Note that RQ1 is broken down into Sections \ref{sec:rq1.1} and \ref{sec:res:info}.

\subsection{RQ1: Submitting VCRs} \label{sec:rq1.1}
As detailed in Section \ref{sec:studes}, we submitted VCRs by form, email, 
or phone-call, as outlined in the \dbrspossessive privacy policy. \\

\begin{figure}[ht]
    \centering
    \includegraphics[width=\columnwidth]{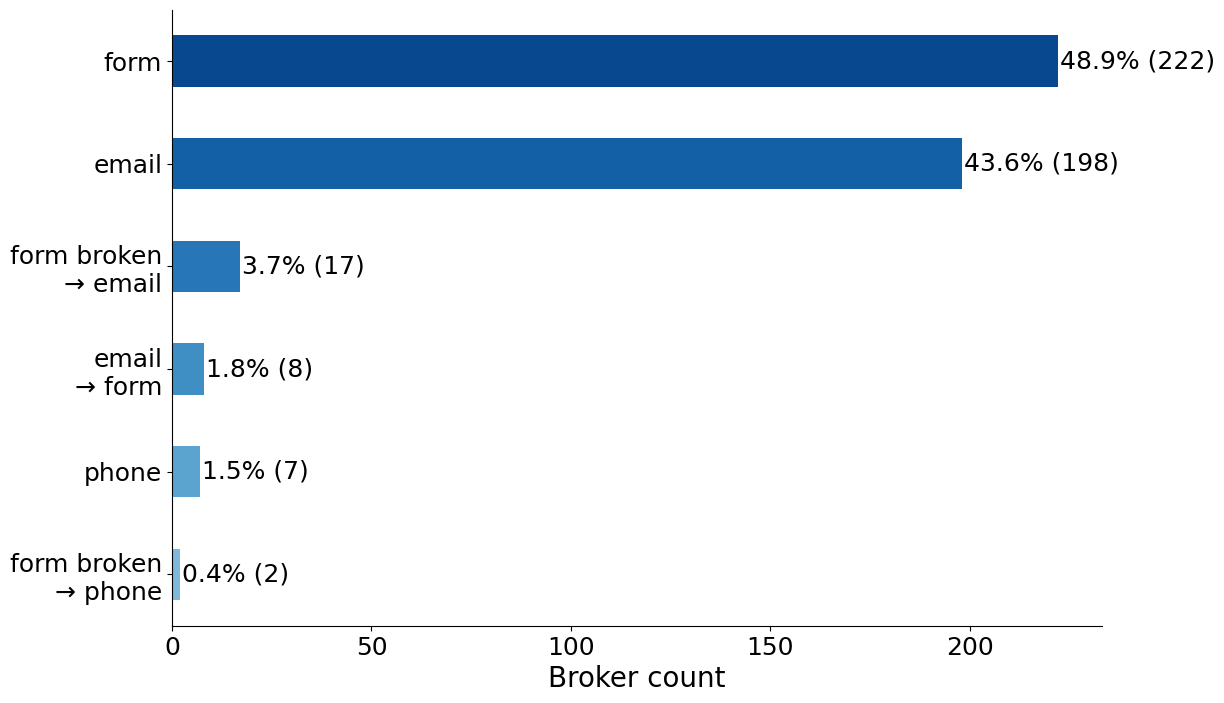}
    \caption{VCR submission methods.}
    \label{fig:vcrcounts}
\end{figure}

\noindent {\bf VCR Submission Methods.}
Figure \ref{fig:vcrcounts} depicts various submission methods encountered and their proportions.
Since there is no CCPA-standardized method to submit VCRs, each \dbr chooses what it prefers, 
which complicates VCR submissions for consumers.
The researcher managed to submit most VCRs (92.5\%) directly by email or by form.

Forms, denoted form-VCRs, were the most common method. While seemingly straightforward 
(just complete and submit), forms presented their own challenges mainly because they are not 
standardized, with each DBR using a different interface and requiring different information. The second common VCR method was email, denoted email-VCR. 

We note that \dbrs can purchase CCPA compliance services from privacy-as-a-service 
companies. OneTrust\cite{onetrust} was the most popular service, used by $56$ \dbrs: 
about $25$\% of form-VCRs were "powered by OneTrust". Curiously, despite using 
OneTrust, form fields varied quite a bit among the $56$ \dbrspossessive form-VCRs.

For $7$ \dbrs, VCRs had to be made by phone, denoted phone-VCRs. This corresponds to 
cases where \dbrs only posted their physical address and phone number, 
leaving no way to contact them electronically.

$27$ submissions involved multiple steps. Upon submitting email-VCRs to 8 \dbrs, 
the researcher received links to resubmit through an online form. This occurred either because 
the \dbr preferred email-initiated requests or because the form was not found in the \dbr's 
privacy policy. Notably, certain \dbrs combine their "request to know" and "request to opt-out/delete" 
forms into one. This practice leads to confusion because the combined forms are frequently 
labeled only as "request to opt-out/delete". We believe that most consumers would not open a form 
titled "delete my data" when attempting to access their data.

$19$ forms could not be submitted, primarily due to broken links in the privacy policy. When such 
forms were encountered, the researcher emailed or called the \dbr.  Some forms failed 
to submit due to an "invalid CAPTCHA", even though no CAPTCHA was displayed on the webpage.
In another irritating case, the form required a frequent shopper ID (confirmed to be a 
string of digits by the \dbr), yet the field only accepted email address formats.

Finally, in a particularly time-consuming request case, the researcher had to call a \dbr 
after being unable to submit the request form, since no email contact was specified in that
\dbr's CCPA privacy policy. After leaving their contact information by voicemail, 
the researcher received an email from the \dbr asking to fill in a form (the link to which was broken) 
or to provide additional information by email. Despite replying with the required details 
and reporting on the form's malfunction, no further response was received. \\

\begin{figure}[ht]
    \centering
    \includegraphics[width=0.9\linewidth]{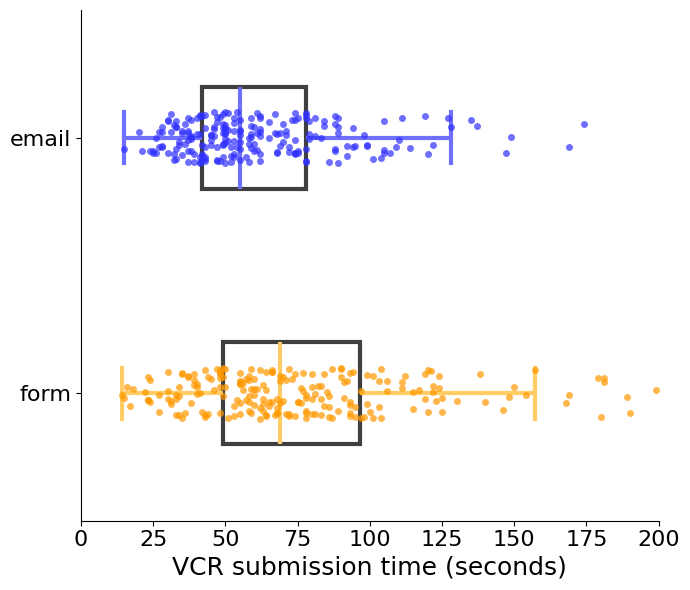}
    \caption{Time needed to submit VCRs.}
    \label{fig:timeForReqs}
\end{figure}

\noindent {\bf VCR Submission Time.}
In total, $9$ hours and $57$ minutes were spent submitting VCRs to all $454$ \dbrs.
This corresponds to the average of $79$ seconds to find the VCR method and to submit a VCR. 
Figure \ref{fig:timeForReqs} shows the time to submit email-VCRs and form-VCRs.
Phone and multi-step procedures are excluded from the Figure.

It took, on average, longer to submit form-VCRs than email-VCRs, with averages of 
$82$ and $66$ seconds, respectively. Shapiro-Wilk tests reject the null-hypotheses that 
email-VCR and form-VCR submission times are normally distributed (p-value $<0.001$), and \new{a 
permutation test} suggests a statistically significant difference in submission times 
(p-value \new{$<0.005$}). Note that, while forms were filled out manually, emails were 
copy-pasted from a template. Submitting OneTrust forms took roughly as long as other forms. 

Submitting phone-VCRs was quite time-consuming. An average phone-call lasted $4$ minutes and 
$37$ seconds. It typically included talking with a company representative or 
leaving a voice message if no one was available. We were surprised by phone representatives 
being totally unaware of CCPA and confused by the nature of our requests.
This was troubling since we made sure to use the phone number explicitly specified
by \dbrs in their privacy policy.\\

\noindent{\textbf {\textsc{Key Takeaways - RQ1 Part 1.}}}
Submitting VCRs is a complex process. While email-VCRs are one of the most commonly used methods, their open-ended nature 
places a burden on consumers, who must properly word their emails to ensure that \dbrs do not simply dismiss 
their requests, and verify that the emailed VCR does not bounce. In comparison, form-VCRs are more 
straightforward, though more time-consuming. Multi-step processes incur an unjustified 
burden since consumers must keep track of (and complete) several steps, which  
are usually due to \dbrs' inadequate privacy policies or outdated contact information. 
Finally, phone-VCRs are plain unhelpful. Even when the calls were answered, employees demonstrated 
an alarming lack of understanding about the nature of the inquiry.


\subsection{RQ1: Identity Verification Process}\label{sec:res:info}
Recall that \dbrs have to verify the consumer identity via VCRs before granting them 
access to their PI. To this end, they ask the consumer to provide a set of PII, 
which is presumably matched with previously collected PII.\\

\noindent {\bf Type of PII Requested}
We first categorize different types of PII requested by \dbrs to perform identity 
verification. Table \ref{tab:InfoReq} shows, for each PII category, the 
number of \dbrs that requested it.

\begin{table}[ht]\caption{Types of PII elements required for VCR submissions.} \label{tab:InfoReq}
\begin{tabular}{lr}
\hline
\textbf{PII type} & \textbf{Count (out of 454)} \\
\hline
Email & 444 \\
Name & 431 \\
Address & Partial:  70, Full: 122 \\
Phone number & 103 \\
Date of birth & Partial: 8, Full: 33 \\
MAID & 26 \\
Official ID & Redacted: 13, Full: 4 \\
SSN & Last 4 digits: 9, Full: 7 \\
Signature/Signed affidavit & 15 \\
LinkedIn URL & 11 \\
Employer & 11 \\
IP address & 8 \\
Multiple-Choice Questionnaire & 5 \\
Selfie & 5 \\
Cookie & 5 \\
Utility bill (redacted) & 6 \\
Previous addresses & 4 \\
Driver's license number & 2 \\
Recently visited locations & 2 \\
Grocery store frequent shopper ID & 1 \\
Gender & 1 \\
Social media usernames & 1 \\
Marital status & 1 \\
\hline
\end{tabular}
\end{table}

Email address and name were requested by over 95\% of \dbrs. Next, home address and phone 
number were requested by 42\% and 23\%, respectively. \dbrs likely use 
home address to confirm California residency and CCPA jurisdiction. 
Sometimes, only partial address information was required, e.g., city, zip code, or state. 
21 \dbrs asked for professional information, e.g., the LinkedIn URL, 
and/or the name of the employer.

32 \dbrs asked for PII that is difficult to retrieve for the average consumer, such as cookie values, 
Mobile Advertising Identifier (MAID), or IP addresses. To obtain cookie information 
on Google Chrome, one needs to \texttt{Inspect} the webpage, click on \texttt{Application}, and find 
relevant cookie values under \texttt{Storage}. On Android devices, user MAID is accessible within 
the phone's settings. However, for iPhone users, Apple does not directly provide access to MAID. 
Instead, iOS users must download a third-party app to view their MAID, which is a bit disconcerting.

\begin{figure}[ht]
\centering
\begin{subfigure}[b]{0.5\textwidth}
   \includegraphics[width=0.9\linewidth]{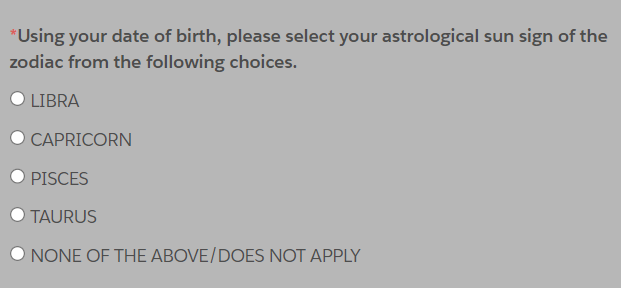}
\end{subfigure}

\begin{subfigure}[b]{0.5\textwidth}
   \includegraphics[width=0.9\linewidth]{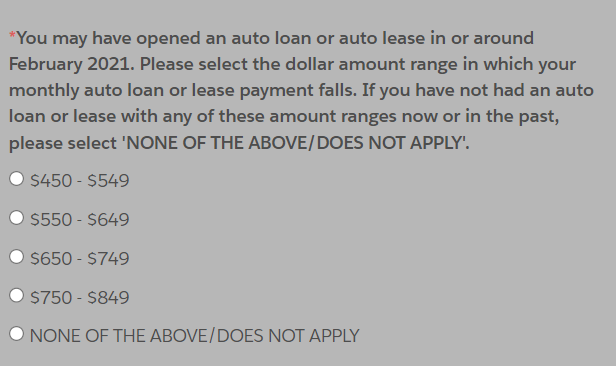}
\end{subfigure}

\caption{Questions from a questionnaire in a form-VCR.} 
\label{fig:questionnaire}
\end{figure}

Five \dbrs requested unusual information: marital status, gender, 
a list of recently visited locations, and a grocery store frequent shopper ID. 
Another five \dbrs presented a multiple-choice questionnaire asking the consumer to 
select their PI from a list. A screenshot of a sample questionnaire is shown in Figure \ref{fig:questionnaire}.

Finally, 45 \dbrs asked for sensitive PII, as defined by CCPA. This included:
full or partial SSN, full or partially redacted copy of a government ID, driver's 
license number, and biometrics, e.g., a signature or a selfie. This includes 4 \dbrs 
that used third-party systems requiring the consumer to take a live selfie 
along with a picture of a government ID. \\

\noindent {\bf Number of PII Elements in the VCR.}
The number of PII elements requested for identity verification varied depending 
on the VCR submission method. Since phone-VCRs were a small minority and mostly unsuccessful, 
a minimal amount of PII was shared through them, usually only phone number and name. 
Most email-VCRs require 2 PII elements, while about half of form-VCRs require 4 or more.
A Mann-Whitney U test indicates that form-VCRs requested significantly more PII 
(p-value $< 0.001$) than email-VCRs. OneTrust forms did not ask for more PII elements 
than other \dbr forms. \\

Submitting email-VCRs involved sharing the researcher's email address, from which the 
VCR was sent, and full name, signed at the end of the email. Seven \dbrs specified 
(in their privacy policy) additional PII elements to include in the email, and 23 
requested additional information after the initial email, in order to continue 
with identity verification. In contrast, form-VCRs typically required all
PII elements up front, at the initial submission time. \\

\noindent {\bf Additional Verification.}
Besides providing PII, a consumer needs to take additional steps to successfully submit form-VCRs. 
For instance, 101 \dbrs required confirming access to the supplied email address 
or phone number through a link or one-time code, or requested a copy of a recent utility bill 
to prove California residence. Many \dbrs with form-VCRs mandated ticking a box to acknowledge 
(under penalty of perjury) that the researcher is the same consumer for whom the request 
was being made. Nevertheless, 13 \dbrs asked for a separate {\em signed affidavit}.

Moreover, 53\% of form-VCRs were CAPTCHA-protected. While CAPTCHAs are used to prevent malicious bot 
activity, CAPTCHA solving is considered difficult for humans. Furthermore, bots are known to be faster
and better than humans at this task \cite{searles2023empirical,bursztein2010good}. Access to 2 
\dbrspossessive form-VCRs resulted in multiple CAPTCHAs: reCAPTCHA and a text-based or math-based CAPTCHA. 
Table \ref{tab:captcha} shows the frequency of CAPTCHA types observed on forms.

OneTrust forms overwhelm consumers with additional verification: they all require CAPTCHA
completion, using either reCAPTCHAs or text-based CAPTCHAs. 77\% require confirming access to the 
supplied email address. In contrast, only 37\% of \emph{non}-OneTrust forms are CAPTCHA-protected. 
Compared to other form-VCRs, OneTrust forms require significantly more additional 
verifications before submission (Mann-Whitney U test p-value $<0.001$). The researcher 
also needed to solve a text-based CAPTCHA every time they wished to access a 
OneTrust portal after VCR submission, e.g., to verify the request's status.\\

\begin{table}[ht]
    \centering
    \caption{CAPTCHA type frequency in CAPTCHA-protected form-VCRs.}
    \label{tab:captcha}
    \begin{adjustbox}{width=0.6\linewidth}
    \begin{tabular}{lr}
    \hline
       CAPTCHA type &  Percentage \\
         \hline
         reCAPTCHA & 72\%\\
         text-based & 20\%\\
         hCAPTCHA & 8\%\\
         math-based & 3\%\\
         \hline
    \end{tabular}
    \end{adjustbox}
\end{table}

\noindent{\textbf{\textsc{Key Takeaways - RQ1 Part 2}}}
The identity verification process for consumers is burdensome and intrusive. While most \dbrs require only a few PII 
elements, the type of these elements varies significantly. Some ask for PII that is inconvenient to obtain, such as 
mobile advertiser IDs or signed affidavits. The process can be highly privacy-invasive, with certain \dbrs requesting 
highly sensitive PII, e.g. SSNs or government IDs. Finally, \dbrs often implement additional verification steps 
(e.g., email confirmation, CAPTCHA solving) before allowing VCR submissions, primarily to prevent spam, yet
further complicating the VCR submission process for consumers.

\subsection{RQ2: $\bm{\mathcal{DBR}}$ responses} \label{sec:resp}
A substantial fraction of \dbrs ($195$ out of $454$) \textbf{never responded} to VCR requests. 
Only $51.5$\% of the \dbrs responded on time, i.e., within 45 calendar days. 
Although no \dbr asked for a permitted extension, $5.3$\% responded after the 
45-day limit, thus failing to comply with the prescribed timeline.


We use the term \textit{responding} \dbrs to denote those that replied to a VCR, 
whether on time or late. The rest,  even if they initially acknowledged the 
VCR or corresponded to get more PII for identity verification, are called {\it non-responding} \dbrs.

\begin{figure}[ht]
    \centering
    \includegraphics[width=1\linewidth]{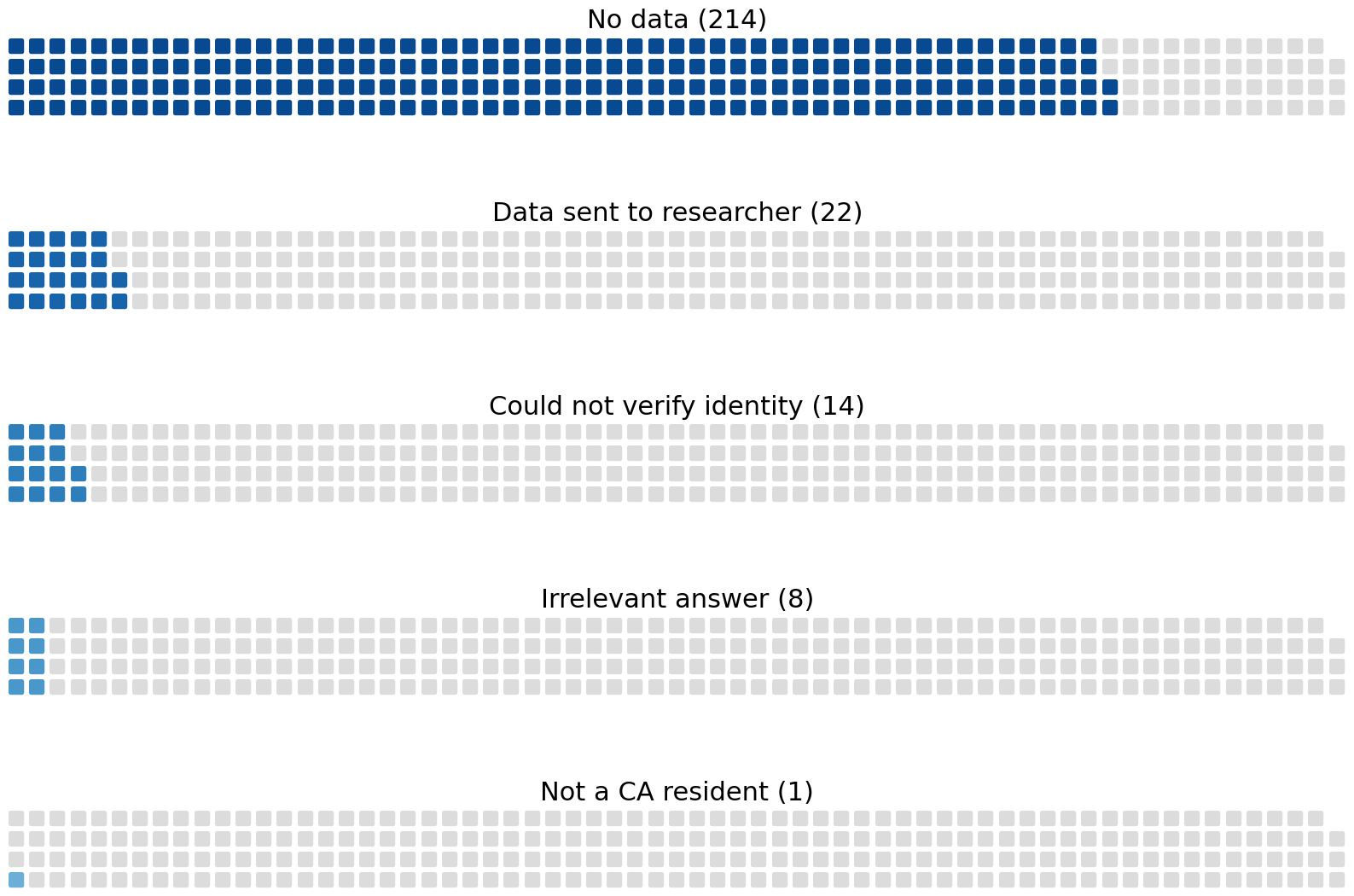}
    \caption{Distribution of answer categories received from \dbrs.}
    \label{fig:brokerAnswers}
\end{figure}

Most responding \dbrs reported having no PI about the researcher. 
Only 22 \dbrs provided some PI, detailed in Section \ref{sec:datacollected}. 
Figure \ref{fig:brokerAnswers} shows various replies from responding \dbrs. 

$14$ \dbrs could not verify the researcher's identity. In $7$ cases, 
their response arrived almost instantly after submitting a VCR form. These \dbrs did not ask 
for additional information to verify the identity. $8$ \dbrs answered with irrelevant information, 
such as stating  that the researcher was now ``opted out'' of future sales or communication. 
Remarkably, one \dbr refused to fulfill the request, falsely claiming that the researcher 
is not a California resident. \\

\begin{figure}[ht]
    \centering
    \includegraphics[width=\columnwidth]{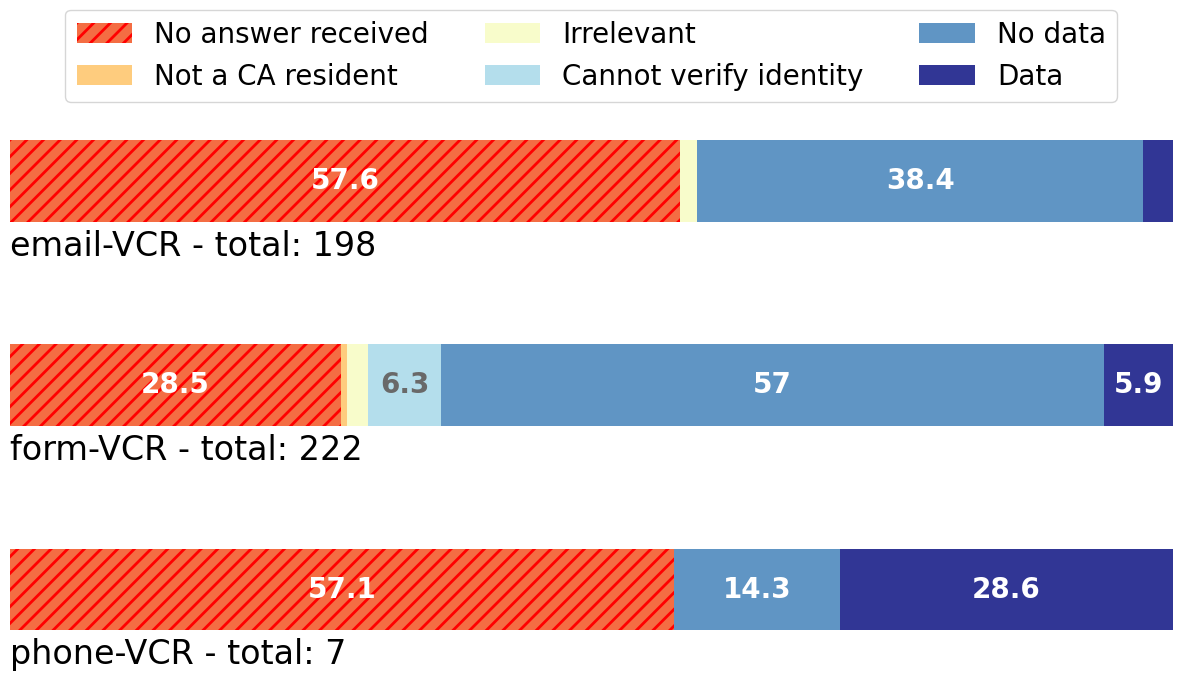}
    \caption{\dbr responses for email, form, and phone VCR submissions (in \%).}
    \label{fig:outputvsmethod}
\end{figure}

\begin{figure}[ht]
    \centering
    \includegraphics[width=\columnwidth]{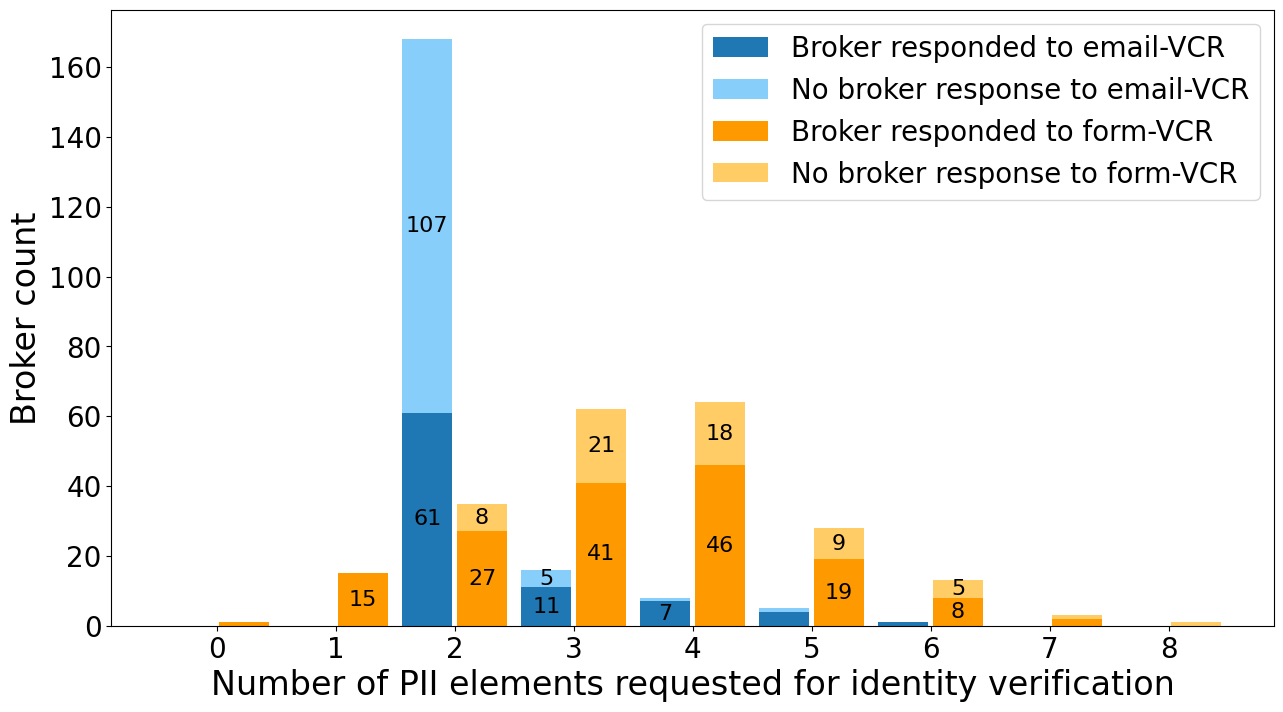}
    \caption{Number of responding \dbrs given increasing number of PII elements requested}
    \label{fig:inforeq}
\end{figure}

\noindent{\bf Response rate.} 
Since a significant fraction of VCRs were unanswered, we examined factors associated with 
receiving a response, such as \dbr location, VCR submission method, number of PII elements requested,
and sensitivity of the PII supplied. Note that the analysis of number of PII elements and sensitivity 
of PII supplied (Figures \ref{fig:outputvsmethod} and \ref{fig:inforeq}) excludes multi-step VCRs.\\
 
\noindent \textit{Geographical location.}
First, we analyzed response rates depending on \dbrspossessive location. We optimistically
expected that \dbrs located in the US, especially those in California, 
would have a better grasp of CCPA and would therefore be more likely to answer. 
$56.6$\% US-based \dbrs and $60$\% non-US \dbrs responded to VCR requests. However, the
latter are geographically dispersed and few in number: 25 total. 

Table \ref{tab:states} provides \dbrspossessive response rate per state, in states where a VCR 
could be submitted to at least 10 \dbrs. Comparing three states with the highest \dbr concentration 
(California, New York, and Florida),  the response rate for California, at $58.1$\%, is only $5.6$ and $1.3$ 
percentage points higher than that of the other two, respectively. In other words, California \dbrs 
perform only marginally better than the US average of 56.6\%.

\begin{table}[ht]
    \centering
        \caption{Response rates of \dbrs in US states.}
    \label{tab:states}
    \begin{tabular}{lcc}
    \hline
     State  &  \dbr Count & Resp. rate\\
      \hline
California       &       93 & 58.1\% \\
New York         &        61 & 52.5\% \\
Florida          &        37 & 56.8\% \\
\hline
Texas            &        23 & 56.5\% \\
Massachusetts    &        20 & 70.0\% \\
Georgia          &        16 & 62.5\% \\
Virginia        &         14 & 50.0\% \\
Colorado         &        14 & 50.0\% \\
New Jersey       &        14 & 71.4\% \\
Illinois          &       12 & 58.3\% \\
Washington      &         11 & 72.7\% \\
Ohio        &        10 & 70.0\% \\
\hline\\
    \end{tabular}
\end{table}

\noindent \textit{VCR submission method.} Beyond geographical factors, we hypothesized that \dbrs 
that used form-VCRs would be more likely to respond to a VCR. The rationale was that \dbrs
with established VCR submission processes would be more inclined to allocate both time and 
personnel to address these requests. Figure \ref{fig:outputvsmethod} shows \dbrspossessive answers 
depending on the VCR submission method. 

Aligning with our assumptions, form-VCRs 
yielded a higher response rate of 71.5\%. Email-VCRs, the second most common VCR submission method, yielded a response rate of 42.4\%. This suggests that the 
majority of these \dbrs \textit{only} meet a minimum regulatory requirement by providing an email address, 
without actually responding to received requests. Finally, phone-VCRs performed worse: only 42.9\% 
received a response. \dbrs consistently failed to return calls after voicemails were left. 
In one notable instance, upon the researcher's request to exercise their CCPA rights, a \DBR 
representative abruptly transferred the call to voicemail. No follow-up communication was received.

Predictably, 19 \dbrs with broken forms that had to be contacted via email or phone exhibited 
low response rates: 41.2\% responded to emails, and none responded to phone calls.
6 of 8 \dbrs that sent a form to complete after an initial email contact 
responded to the VCR, and 2 did not answer after they specifically pointed to a form. \\

\noindent \textit{Number of PII elements requested.} 
To verify the requester's identity, \dbrs ask for PII to match with their own records. 
As confirmed in Section \ref{sec:res:info}, due to the lack of standards in CCPA legislation 
each \dbr asks for different types of PII. 

Figure \ref{fig:inforeq} shows the response rate for email-VCRs and form-VCRs depending on the 
number of PII elements requested for identity verification. 
168 email-VCRs were given 2 PII 
elements by the researcher: email address and name, and were completed in 36.3\% of cases only. 
30 email-VCRs that required more than 2 PII elements were answered at a higher rate (76.7\%). 
The number of PII elements requested in form-VCRs had no effect on the response rate. \\

\noindent \textit{Sensitive PII requested.}
Finally, we checked whether \dbrs that asked for sensitive PII responded to VCRs. 
Following CCPA's definition of "sensitive personal information", Table \ref{tab:sensitiveRespRate} 
lists the seven elements of sensitive PII occasionally requested, and \dbrspossessive response rate when
requiring these sensitive PII elements. 
Most \dbrs that required a full or redacted copy of a 
government ID completed the VCR. Meanwhile, only 2 of 6 \dbrs asking for a full SSN responded. 
This result is especially concerning: 
\begin{quote}
{\sf \textit{Who, now, has access to this information?}}
\emph{Does exercising one's CCPA rights in turn put a consumer at risk of identity theft?}
\end{quote} 
Two \dbrs that requested the last 4 digits of the researcher's SSN or that were sent a 
redacted copy of a government ID responded saying that they 
could not verify the researcher's identity: sharing sensitive PII was ineffective in these cases. \\

\begin{table}[]
    \centering
    \caption{\dbrs response rate when sensitive PII is asked.}
    \label{tab:sensitiveRespRate}
    \begin{tabular}{lr}
    \hline
       Sensitive PII & Completed VCRs \\
       \hline
       Last 4 digits of SSN  & 5/9 (55\%) \\
       Full SSN & 2/6 (33\%)\\
       Redacted ID & 11/13 (85\%) \\
       Full ID  & 3/4 (75\%)\\
       Driver's License Number & 1/2 (50\%) \\
       Signature/signed affidavit & 10/15 (67\%)\\
       Selfie & 4/5 (80\%)\\
         \hline 
         Total & 29/45 (64\%) \\
         \hline
    \end{tabular}
\end{table}

\begin{figure}[ht]
\centering
\begin{subfigure}[b]{0.5\textwidth}
   \includegraphics[width=0.9\linewidth]{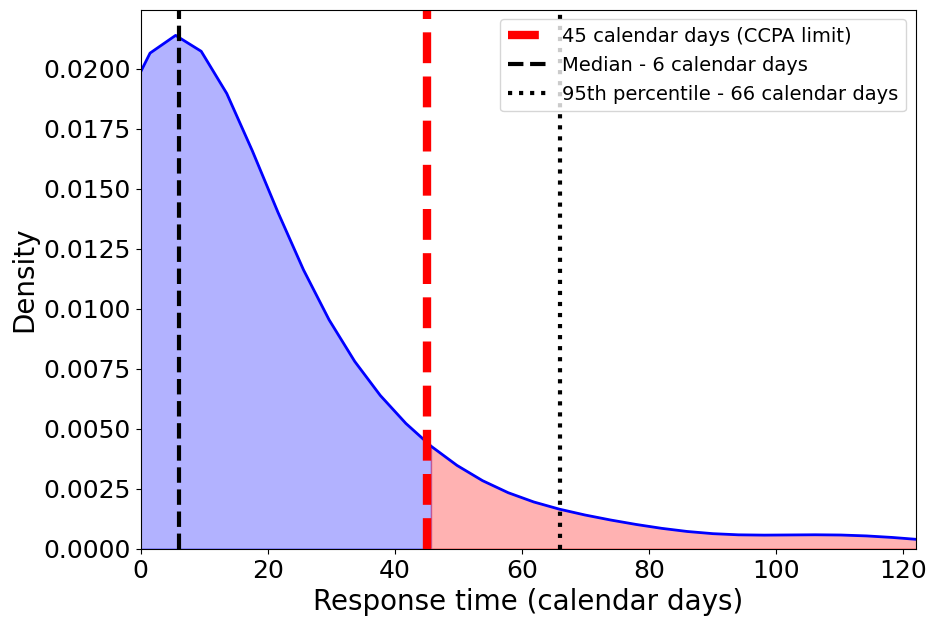}
   \caption{\dbr response time}
    \label{fig:density}
\end{subfigure}

\begin{subfigure}[b]{0.5\textwidth}
   \includegraphics[width=0.9\linewidth]{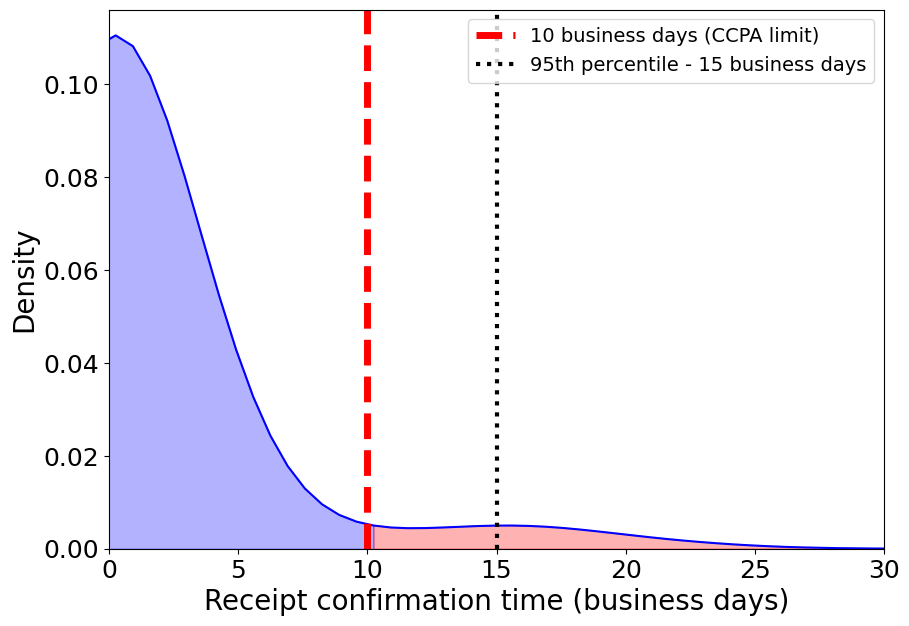}
   \caption{\dbr VCR receipt confirmation time}
    \label{fig:confirm}
\end{subfigure}

\caption{Distribution of VCR completion and receipt confirmation time.} \label{fig:densitytimes}
\end{figure}

\noindent {\bf Response Timeline.}
Figure \ref{fig:density} shows the distribution of the response timeline, in calendar days, 
between VCR submission and \dbr response, for the 259 responding \dbrs. Recall that CCPA requires
\dbrs to respond within 45 calendar days.
Half of the responding \dbrs answered the VCR within the first 6 days, and 90.7\% responded within the allotted 
45 days. \dbrs also have to confirm receipt of a VCR within 10 business days. We considered that every 
form-VCR was immediately confirmed as received, through the form submission confirmation message. 
For all other VCRs, we assume that the \dbr's first email to the researcher was confirmation of VCR receipt. 
Figure \ref{fig:confirm} shows the distribution of time, in business days, between VCR submission and 
confirmation of receipt. 93.6\% of responding \dbrs confirmed receipt of the VCR within the allotted 
10 business days. In fact, 81.8\% confirmed receipt of the VCR within one business day,  typically via 
an automatic reply system. Of the 195 \dbrs that did \emph{not} respond, 69 confirmed receipt of 
the VCR without following up. \\

\begin{figure}[ht]
    \centering
    \includegraphics[width=0.9\columnwidth]{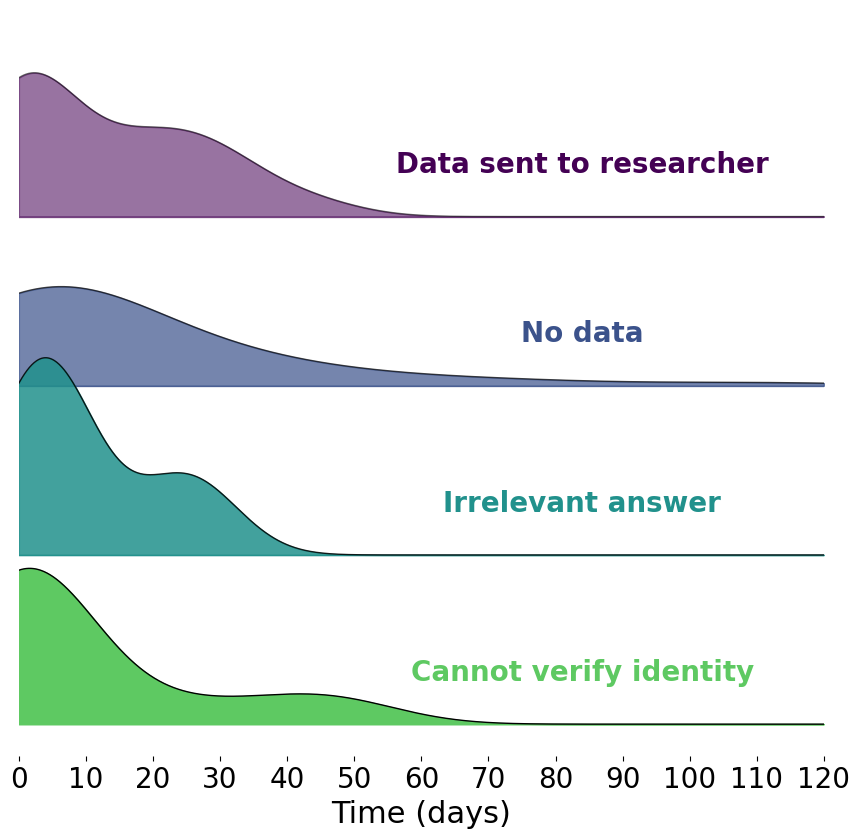}
    \caption{\dbr response time depending on answer content.}
    \label{fig:timeVSanswer}
\end{figure}

\noindent \textbf{Response contents.}
Figure \ref{fig:timeVSanswer} shows the response distribution of response times categorized 
by the type of answer provided. 
\dbrs that had PI about the researcher took longer to answer: this could be due to either 
a more thorough identity verification process, or the effort to find the collected PI.
On the other hand, \dbrs that could not verify the researcher's identity or replied with 
irrelevant data (about opt-out or deletion rights) answered more expeditiously. 
However, it is uncertain whether their answers can be trusted, e.g., did they reply
just to quickly "complete" (i.e., get rid of) the VCR?\\

\noindent \textbf{OneTrust.}
OneTrust form-VCRs yielded a response rate of 79\%, notably higher than the 69\% rate of 
non-OneTrust form-VCRs. Therefore, while OneTrust form-VCRs required additional 
verifications (CAPTCHA, email confirmations), they had a higher "success" rate.
We conclude that the use of such an automated system, preferably standardized, 
might help \dbrs meet compliance requirements.\\

\noindent \textbf{\textsc{Key Takeaways - RQ2.}}
Measuring response rate revealed widespread non-compliance.
A significant fraction of \dbrs failed to respond to VCRs, with only about half answering 
within the prescribed 45-day timeframe. Specifically, email and phone-VCRs were answered only in 
42\% of cases. Form-VCRs performed better, with a 71\% response rate, potentially indicating that 
providing a form demonstrates a stronger commitment to CCPA compliance. Alarmingly, even \dbrs 
requesting PII exhibited low response rates. 

Among responding \dbrs, most adhered to CCPA-mandated 45-day limit. The pattern exposed a stark 
divide: \dbrs tend to either fully comply with or completely disregard CCPA obligations. 
Overall, these findings paint a rather disappointing picture of CCPA compliance.

\subsection{RQ3: PI received from $\bm{\mathcal{DBR}\mathsf{s}}$}\label{sec:datacollected}
We analyze received PI to evaluate accessibility, accuracy, PI leakage, and tangentially, 
mapping PI requested in VCRs to the sensitivity of information \dbrs revealed. Recall that $22$ 
\dbrs provided PI about the researcher. These \dbrspossessive registered addresses span 11 
states within the US, and three countries (USA, Germany, Canada). Table \ref{brokdata} 
presents details regarding the VCR process and the PI received from these \dbrs.
The following PI was obtained: 
\begin{compactitem}
    \item \textit{Confidential PI (2).} A credit reporting agency sent a 
    credit report, and a major data analytics company sent past and current car insurance 
    information: policy numbers, dates, coverage, and listed drivers. This information is very
    sensitive.
    
    \item \textit{Behavior inference data (3).} These \dbrs provided such data for advertising purposes, e.g., 
    {\tt "Samba TV $\rightarrow$ Ad Exposure $\rightarrow$ Electronics"}.
    
    \item \textit{Publicly available professional information (6).} These \dbrs provided a copy of the 
    researcher's public LinkedIn profile. 

    \item \textit{Physical addresses and cookie values (2).} These \dbrs provided a limited amount 
    of PI, only giving one or two home addresses and some website-specific cookie ID. 
    
    \item \textit{Data from the VCR (5).} These provided information that they likely just obtained from
    the VCR itself. These include only the researcher's name and email address, and VCR-provided IP 
    address, cookie ID, and mobile advertiser ID.
    
    \item \textit{Categories of collected PI (3).} These provided the categories, not the actual PI. 
    Thus, we cannot verify whether these \dbrs actually collected the researcher's PI or if a 
    generic answer was given.\\
    
\end{compactitem}

\begin{table*}[]
\caption{PI Received from \dbrs in Response to VCRs.} \label{brokdata}
\resizebox{\textwidth}{!}{
\begin{tabular}{llllrrl}
\hline
 & \textbf{Type of PI received} & \textbf{PI transmission method} & \textbf{File format}  
 & \textbf{\# PII for ID verif.} & \textbf{Ans. time (days)} & \textbf{Location} \\
 \hline
 1 & Credit report & Postal mail & N/A & 6*  & 0 & Georgia \\
2 & Car insurance information & Link to PI sent via postal mail & pdf & 7*  & 1 & Georgia \\
\hline
3 & \multirow{3}{*}{Behavior inference for advertising} & Email attachment & pdf  & 4 & 44 & California \\
4 &  & Expiring link & xlsx & 4  & 29 & Virginia \\
5 &  & Directly website accessible & csv  & 0  & 0 & New York \\
\hline
6 & \multirow{6}{*}{LinkedIn profile data} & Email attachment & pdf  & 2 & 0 & California \\
7 &  & Email attachment & pdf  & 8  & 1 & New York \\
8 &  & Email attachment & pdf & 3  & 5 & Washington \\
9 &  & Email attachment & json  & 2  & 17 & Washington \\
10 &  & Password-protected email attachment
& xlsx  & 5 & 26 & Delaware \\
11 &  & Email content & N/A  & 4  & 1 & California \\

\hline

12 & Physical addresses and company-specific cookies & Expiring link & csv & 7*  & 12 & California \\
13 & Website cookie ID & Directly website accessible & N/A  & 2  & 14 & Canada \\
\hline
14 & Email address & Expiring link & html  & 4  & 34 & Virginia \\
15 & Shortened first name & Directly website accessible & N/A & 3 & 1 & Colorado \\
16 & IP address information & Email attachement & json & 3 & 4 & Nevada \\
17 & Cookie ID and mobile advertiser ID & OneTrust Portal & N/A & 5  & 32 & Germany \\
18 & Name, email address & OneTrust Portal & pdf  & 4*  & 22 & New York \\
\hline
19 & \multirow{3}{*}{Categories of PI} & Phone-call & N/A  & 2  & 0 & Utah \\
20 &  & OneTrust Portal & txt  & 4& 0 & Texas \\
21 &  & Email attachment & xlsx, json & 5 & 24 & Massachusetts \\
\hline
22 & History/details of services provided by the company &  OneTrust Portal & pdf & 5 & 20 & Georgia\\
\hline
\end{tabular}}
\begin{flushright}
{\footnotesize * indicates sensitive PII was requested}
\end{flushright}
\end{table*}

\noindent{\bf PI Delivery.}
Two \dbrs that sent the most confidential PI did so by postal mail: the PI itself or a printed download link 
to the PI was sent to the researcher's home address. This measure ensured that only the consumer with the 
address matching that in the \dbrspossessive databases would receive the PI. It maintains data security as long 
as \dbrs are confident in the accuracy of their databases. Incidentally, the same two \dbrs required a 
higher-than-average number of PII elements and sensitive PII for identity verification. 

Other \dbrs sent the PI via email or OneTrust portals. Three had PI directly accessible through their website,
e.g., one prompted the researcher to press a "show me the data" button, after which they could download a 
CSV file containing behavior inference data. \dbrs usually provided the PI as PDF, CSV, or XLSX files. 
One \dbr sent files as a series of nested HTML files with non-descriptive titles, making sense of which
is likely to be unintuitive for the average consumer. \\

\noindent{\bf Accuracy of PI.}
Behavior inference data received from 4 \dbrs was vague, with multiple age ranges or 
contradictory interests. Some \dbrs sent thousands of data points, and others dozens. 
Certain inferences were inaccurate, e.g., two \dbrs provided data inferring that the researcher 
is Hispanic or Spanish-speaking, which is not the case.\\

\noindent{\bf PI Leakage.}
Alarmingly, the \dbr from which the researcher received car insurance information included 
sensitive PI about the following consumers:
\begin{compactenum}
    \item The researcher's family members, included as drivers in the researcher's policy: this 
    might be acceptable, since the researcher entered this information themselves when starting the policy.
    \item The researcher's housemate, on whose policy the researcher was listed as an excluded driver 
    (an individual within the same household specifically listed as not covered under the car insurance policy).
    We consider this to be a privacy leak. The researcher was sent a third-person's PII, of which the researcher 
    had no prior knowledge.
    
    Specifically, the researcher received the policy number, start and end dates, coverage limits, and 
    \textbf{7 out of 8 characters} of the housemate's driver's license number. While the first 
    half of the report redacted the first 5 characters of the 8-character-long California driver's 
    license number (e.g. XXXXX678), the second half of the report redacted the last 4 characters (e.g. A123XXXX). 
    Accordingly, the researcher obtained 7-out-of-8 characters (e.g. A123X678) and would only need to
    guess the 5th character, a decimal digit, to obtain their housemate's full driver's license number.
    We note that CCPA prohibits disclosing any consumer's "...driver's license number or other government 
    issued identification number [in response to a request to know]" (Tit. 11 Div. 6 §7024(d)) \cite{CCR}.\\
\end{compactenum}

\noindent \textbf{\textsc{Key Takeaways - RQ3.}}
While a few \dbrs shared confidential PI or advertising behavioral inference data, the majority of PI shared 
was not particularly meaningful. Most \dbrs simply returned the PII provided by the researcher, 
or at best, copied LinkedIn profile data.

\dbrs typically transmitted PI via email or OneTrust portal. Notably, two \dbrs that shared the most sensitive 
PII opted for postal mail delivery, demonstrating enhanced security measures. This suggests that \dbrs 
tailored their delivery methods to the sensitivity of the PII transmitted. File formats, with only one exception, 
were easily accessible and transmittable, following CCPA requirements.
Surprisingly, one broker leaked sensitive PII about a third party.


\section{Discussion}\label{sec:disc}
Given the study results, we now consider privacy tradeoffs in the VCR process, 
unexpected and noteworthy incidents, and finally recommendations stemming from these, 
for consumers, data brokers, and policymakers. 

\subsection{The Privacy Paradox}
The goal of privacy laws is to give consumers greater control over their PI. Ironically, exercising one's 
CCPA-given privacy rights introduce new privacy risks and vulnerabilities. The researcher sent a 
large amount of PII to hundreds of \dbrs, 95\% of which either did not respond or did not previously 
collect any PI about the researcher. In one email informing that no PI was collected about the researcher, 
a \dbr even stated, about the email address and name given to submit the VCR: 
\begin{quote}
"In fact, we never had this information until receiving it in your email below." 
\end{quote}
This highlights current privacy issues stemming from exercising one's CCPA rights.
It is especially concerning when \dbrs request sensitive PII in VCRs. As 
noted in Section \ref{sec:resp}, 6 \dbrs asked for the researcher's full SSN, 
and 4 of them did not respond to the VCR. This raises serious concerns about 
identity theft risks created by the VCR process, which is intended to protect consumer privacy. 
Not only does the consumer submitting the request become vulnerable, but privacy 
threats to other individuals also emerge, i.e., as mentioned earlier,
the researcher received sensitive PI about another consumer. On a related note, prior work 
shows that an impersonator could easily receive another consumer's PI, 
since the PII required for identity verification is often public or easily 
obtainable \cite{pavur2019gdparrrrr,DBLP:conf/icws/BufalieriMMS20, DBLP:conf/soups/MartinoRWQLA19}.

\noindent This leaves two imperfect choices:
\begin{compactenum}
    \item \dbrs should require a copious amount of (potentially sensitive) PII to more accurately 
    identify the consumer, which is detrimental to consumer privacy.
    \subitem OR
    \item \dbrs should require less PII, thus favoring consumer privacy. However, 
    they would be more prone to data breaches due to ease of impersonation resulting from a
    simplified identity verification process.
\end{compactenum}

\subsection{Unexpected and unintended outcomes}
Even though the researcher only asked to exercise their PI access rights, many \dbrs, 
along with responding to VCRs, announced that they were putting the researcher's information 
on their opt-out list. They either assume that the researcher cares about their privacy and 
anticipate an opt-out/deletion request, or they hope to not receive any more VCRs 
from the researcher by proactively adding them to their opt-out list.

Several factors may explain why such a small percentage of \dbrs shared collected PI with the researcher. 
Recall that only 22 \dbrs, 4.6\% of all contacted \dbrs, and 8.1\% of all responding \dbrs, 
shared PI which they collected about the researcher. Beyond non-compliance, there are several potential reasons.
    CCPA only requires \dbrs to share PI that was collected within 
    the past 12 months (1798.130(a)(2)(B))\cite{CCPA}. If PI was collected earlier, \dbrs might
    not have to disclose it. In addition, if the collected PI is considered "public" in any way, it is 
    not considered PI under CCPA, and \dbrs do not have to disclose it (1798.140(v)(2))\cite{CCPA}.

Finally, we had no way of verifying the veracity of responding \dbrs that claimed they had no PI.

\subsection{Noteworthy incidents}
The study revealed several noteworthy incidents: \\

\noindent \textbf{Noncompliance.}
Four \dbrs added the researcher's email address to their marketing/newsletter list, 
which clearly violates CCPA (1798.130(a)(7)) \cite{CCPA}. Two of three major US credit 
reporting agencies (Experian, Equifax, TransUnion) did not respond to our request. One \dbr 
that definitely had the researcher's exact current address (it was one of the multiple-choice 
questionnaire options) did not answer the VCR. \\

\noindent \textbf{Inaccessible VCRs.}
One \dbr required the researcher to provide cookie values from a specified website. The website 
was down. Although we communicated this to the \dbr, no reply was received.
One \dbr required email access confirmation by clicking a link in an email, even
though no link was included in that email.\\

\noindent \textbf{Oddities.}
A couple of \dbrs used email obfuscation in their privacy policy, making it harder for both 
bots and humans to send emails. We also observed that employees of some \dbrs visited the 
researcher's LinkedIn profile, probably for identity verification purposes.

\subsection{Recommendations} 
We now make a few improvement recommendations for all three stakeholders involved: policymakers, \dbrs, and consumers.\\

\noindent\textbf{Policymakers.} 
Policymakers need to mandate standardization of both the VCR submission process and the maximal set of information 
that \dbrs should be allowed to request for identity verification purposes. More importantly, increased enforcement 
is needed through random audits \new{and legal actions against non-compliant \dbrs. While CPPA already 
fined some \dbrs for failing to register \cite{CPPA_2025,CPPA_2025c,CPPA_2025b}, we are unaware of any \dbrs facing legal repercussions for failing to process consumer requests.} 

Plus, there needs to be a standardized process for consumers to lodge grievances against 
unresponsive or otherwise non-compliant \dbrs. Finally, consumers need to be informed about a history of such grievances
against each \dbr (e.g., why bother submitting yet one more VCR to a \dbr that has numerous recent complaints).\\

\noindent{$\bm{\mathcal{DBR}\mathsf{s}.}$} 
Of course, we strongly recommend that \dbrs must comply with CCPA. We also suggest that \dbrs provide direct links 
to relevant VCR forms or email contacts as part of their CPPA registration, instead of providing a generic link to 
their privacy policy. This may help \dbrs to not confuse different rights: \dbrs should not opt-out consumers from 
the sale of their data if such a request was not made.

Furthermore, \dbrs must make the VCR submission process easy for consumers. If asking for online identifiers is 
essential (e.g., via cookie ID or MAID), obtaining such information should be trivial.
Finally, \dbrs must train designated employees to properly handle responding to phone-VCRs and email-VCRs. \\

\noindent\textbf{Consumers.}
Consumers should exercise caution when submitting VCRs and, if problems are encountered, file official grievances 
using the CPPA complaint form \cite{cppa_complaint_form}.

\subsection{Limitations}
Throughout this study, one researcher sent VCRs to \dbrs. We acknowledge that having multiple 
individuals submit VCRs would yield more data. This is the subject of future work.
In a similar vein, studying CCPA compliance for non-existent (or deceased) individuals might
yield a better overall picture. However, that would also trigger more ethical issues.
Comparing \dbrspossessive compliance with other privacy laws, e.g., GDPR, would also be interesting. 
However, GDPR unfortunately does not mandate \dbr registration.

Our findings likely underestimate the full extent of non-compliance in the \dbr ecosystem, since
our methodology only assessed entities that have already fulfilled the basic legal 
CCPA obligation of \dbr registration. Unregistered data brokers (which by definition are already 
failing to comply with CCPA) were beyond the scope of this study.


\new{Our results, in particular, the number of \dbrs that had data about the 
researcher and their responses, might have been influenced by the specific researcher's
characteristics and online behavior. The researcher in question is under 30 years of
age, privacy-conscious, and not very active on social media platforms.
This naturally limited the amount of data collected by \dbrs. It is likely that
\dbrs would have more data on older individuals, and/or those with richer 
financial histories (e.g., loans, debts, real estate, etc.), as well as ones 
who are more active on social media.}

\section{Conclusion}\label{sec:concl}
This comprehensive study of 543 California-registered data brokers reveals concerning patterns of
CCPA (non-) compliance. A substantial portion of data brokers (43\%) completely ignored VCRs. 
This widespread non-compliance undermines CCPA's efficacy and highlights significant enforcement gaps in current privacy regulation. 
Our findings expose a troubling paradox: exercising privacy rights under CCPA introduces new privacy vulnerabilities. 
Consumers must provide personal information -- sometimes including sensitive PII, such as SSNs, government IDs, and biometric data -- to data brokers who may never respond. 
This creates a privacy catch-22 where consumers must risk exposing additional personal data to potentially untrustworthy entities to learn what information these brokers already (do not) possess. 
The lack of standardization in verification procedures, places the burden on consumers to navigate each broker's unique process, providing various levels of personally identifiable information and creating substantial barriers to accessing their own data. 

\section*{Acknowledgments}

We thank IEEE S\&P reviewers for their constructive feedback. This work was supported in part by funding from NSF Award SATC 1956393,  
and a 2024 Guggenheim Fellowship.

%% file: appendix.tex
\appendices

\section{Email template}\label{app:email}

Subject: 
Request for Information Under CCPA “Right to Know/Access”\\

\noindent To whom it may concern,\\

\noindent I am writing to you in order to exercise my rights under the California Consumer Privacy Act (CCPA) "right to know". \\

\noindent Please kindly provide the following information pertaining to my personal data:
\begin{itemize}
\item The categories of personal information collected \textbf{and} a copy of all specific pieces of information collected about me.
\item The categories of sources of my personal information.
\item The purposes for collecting my personal information.
\item The categories of third parties with whom my personal information is shared or sold.
\item The categories of my personal information that was sold or shared to third parties.
\end{itemize}

\noindent Please let me know if you need any further information from me.\\

\noindent Thank you for your prompt reply.\\

\noindent Sincerely,\\

\noindent <Name>

\newpage
\section{Meta-Review}

\subsection{Summary}
This paper studies the compliance of data brokers with the California Comsumer Privacy Act (CCPA). One of the researchers manually submitted VCRs (Verifiable Consumer Requests) to all 543 data brokers that were registered with the California data broker registry in 2024. The paper analyzes the process of submitting these VCRs and the responses of these data brokers and aims to identify any potential non-compliance regarding the CCPA. It finds that most data brokers are not compliant, and with a few engaging in explicitly illegal behavior.

\subsection{Scientific Contributions}
\begin{itemize}
\item Independent Confirmation of Important Results with Limited Prior Research
\item Provides a Valuable Step Forward in an Established Field
\end{itemize}

\subsection{Reasons for Acceptance}
\begin{enumerate}
\item The paper provides independent confirmation of important results with limited prior research. The paper conducts a large-scale study of data brokers and their compliance; this has been understudied in prior work. In addition to presenting the results of its requests, the paper provides a detailed overview of the challenges consumers might encounter while attempting to submit a VCR.
\item The paper provides a valuable step forward in an established field. While there has been prior work on data brokers, this paper takes an exhaustive approach towards evaluating data brokers' compliance with existing regulations and surfaces important findings.
\end{enumerate}

\subsection{Noteworthy Concerns} 
The paper is about one researcher's experience with submitting VCRs and the results (i.e., the data that was shared or not shared) are limited by the data that specific companies have about the researcher. The findings would be stronger if multiple researchers submitted requests or if the authors relied on data donations from participants.